\documentclass[twocolumn,pra,aps,showpacs,superscriptaddress]{revtex4-1}
\usepackage[utf8]{inputenc}
\usepackage{url,psfrag,graphicx}
\usepackage{dcolumn}
\usepackage{amsmath,amssymb}
\usepackage{pstricks}
\usepackage{epsfig,placeins,float}
\usepackage{hyperref}

\usepackage{soul}

\def\<{\left<}
\def\>{\right>}
\def\ket|#1>{\left|#1\right>}
\def\bra<#1|{\left<#1\right|}
\def\elem<#1|#2|#3>{\left<#1\right|#2\left|#3\right>}
\def\({\left(}
\def\){\right)}

\def\R{{\mathbb R}}

\def\pmat#1{\begin{pmatrix}#1\end{pmatrix}}
\def\{{\left\lbrace}
\def\}{\right\rbrace}
\def\beq{\begin{equation}}
\def\eeq{\end{equation}}

\def\max{\mathrm{max}}
\def\Tr{\mathrm{Tr}}

\def\eff{\mathrm{eff}}

\makeatletter
\def\inmod#1{\allowbreak\mkern5mu({\operator@font mod}\,\,#1)}
\makeatother

\begin{document}

\title[Short Title]{Engineering large end-to-end correlations 
	in finite fermionic chains}

\author{Hern\'{a}n Santos}
\affiliation{Dep. de F\'{\i}sica Fundamental, Universidad Nacional de
	Educaci\'on a Distancia (UNED), Madrid, Spain}
\affiliation{Dep. de F\'{\i}sica de la Materia Condensada, Universidad Aut\'onoma de Madrid, Cantoblanco, 28049 Madrid, Spain}

\author{Jos\'{e} E. Alvarellos}
\affiliation{Dep. de F\'{\i}sica Fundamental, Universidad Nacional de  Educaci\'on a Distancia (UNED), Madrid, Spain}

\author{Javier Rodr\'{i}guez-Laguna}
\affiliation{Dep. de F\'{\i}sica Fundamental, Universidad Nacional de  Educaci\'on a Distancia (UNED), Madrid, Spain}

\date{September 15, 2018}

\begin{abstract}
We explore deformations of finite chains of independent fermions which
give rise to large correlations between their extremes. After a
detailed study of the Su-Schrieffer-Heeger (SSH) model, the trade-off
curve between end-to-end correlations and the energy gap of the chains
is obtained using machine-learning techniques, paying special
attention to the scaling behavior with the chain length. We find that
edge-dimerized chains, where the second and penultimate hoppings are
reinforced, are very often close to the optimal configurations. Our
results allow us to conjecture that, given a fixed gap, the maximal
attainable correlation falls exponentially with the system size. Study
of the entanglement entropy and contour of the optimal configurations
suggest that the bulk entanglement pattern is minimally modified from
the clean case.
\end{abstract}


\maketitle


\section{Introduction}

Two fundamental elements in quantum many-body physics are strong
correlations and entanglement \cite{Amico.08}, which constitute a
basic resource in quantum communications and quantum computation
\cite{Nielsen.00} and a key component of most quantum technologies.
Moreover, the ground states (GS) of quantum systems are known to
present very interesting entanglement properties \cite{Eisert.10},
such as the {\em area-law:} for a gapped system, the entanglement
entropy of a certain block is typically proportional to the block
boundary \cite{Sredniki}. Gapless systems, on the other hand, usually
present logarithmic corrections which can be assessed making use of
conformal invariance \cite{Vidal.03}.

Some (gapless) deformed systems can violate maximally the area law and
present volumetric entanglement, for example the so-called {\em
  rainbow state} \cite{Vitagliano.10,Ramirez.14,Ramirez.15}, a {\em
  valence bond solid} (VBS) where the fermionic bonds are
concentrically placed around the center. It can be built as the ground
state of an open chain of free fermions, whose hoppings decay
exponentially from the center. Nonetheless, the energy gap for the
rainbow state decays exponentially with the system size, thus making
it difficult to implement in actual quantum devices.

The aim of this article is to determine deformations of a quantum
independent fermionic chain which attain a maximal correlation between
its extremes, while keeping an appreciable energy gap. Note that,
according to Hastings' theorem, one-dimensional (1D) gapped systems
must fulfill the area law \cite{Hastings}, proving that it is
impossible to obtain a rainbow state as the GS of a 1D gapped
Hamiltonian. Nonetheless, large end-to-end correlations on a gapped
system are not explicitly forbidden.

\begin{figure*}
\includegraphics[width=12cm]{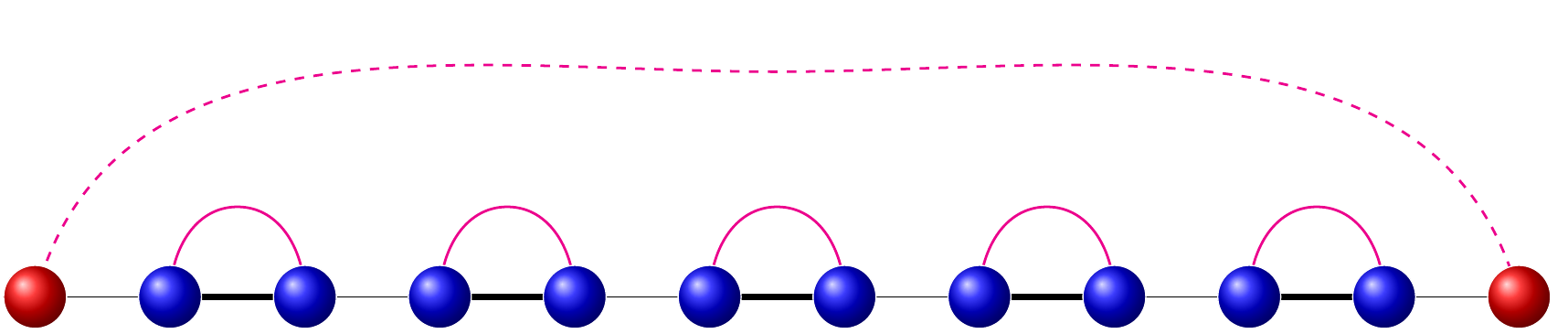}
\caption{An open dimerized fermionic chain with alternate lighter
  (thinner black lines) and stronger links (thicker black lines). The
  stronger links will tend to form a valence bond (red line). A large
  distance correlation between the first and last sites may appear
  (indicated as a dashed red line).}
\label{fig:illust_1}
\end{figure*}

Our study begins with the well-known Su-Schrieffer-Heeger (SSH) model
of a dimerized fermionic chain \cite{Su.79,Heeger.88,Sirker.14}, whose
links alternate between a weak and a strong value, which constitutes a
paradigm for topological insulators \cite{Asboth}. When the first and
last links are weak, an edge state can appear in the form of a valence
bond between the first and last sites, thus inducing a large
correlation between them, see Fig. \ref{fig:illust_1} for an
illustration. Unfortunately, the energy gap required to excite away
this edge state decays too fast with the system size. Yet, it will
provide the essential clues to explain the optimal deformations.

Next, we developed a machine-learning algorithm to obtain the
deformations which maximize the end-to-end correlation (in absolute
value) for a fixed chain length and energy gap. We show that, in many
cases, these configurations are {\em edge-dimerized} chains, where the
second and last links are reinforced. We show that the maximal
correlation obtained with our algorithm decays exponentially with the
system size and with the energy gap. We would like to emphasize that
this article only provides a proof-of-principle strategy to establish
large correlations among distant sites of a quantum system while
retaining a large enough gap. Thus, our conclusions regarding the
scaling behavior of the maximal correlation remain tentative and need
further work.

This article is organized as follows. Sec.\ref{sec:model} presents the
model employed, independent lattice fermions. Dimerized open chains
are discussed in Sec.~\ref{sec:dimerized}. The machine-learning
procedure is described in Sec.~\ref{sec:engineering}, along with the
results obtained for the optimal correlation. This leads to the study
of edge-dimerized chains in Sec.~\ref{sec:modulated}. Our conclusions
are summarized in the last Section.


\section{Model}
\label{sec:model}

Let us consider a chain of $L$ sites where $N_e$ independent spinless
Dirac fermions move. An inhomogeneous tight-binding Hamiltonian can be
written in the following way:

\beq
H=-\sum_i t_i \ c^\dagger_i c_{i+1} + h.c.
\label{eq:ham}
\eeq
where $c^\dagger_i$ is the creation operator at site $i$ and the $t_i$
are the local hopping amplitudes. We will consider $N_e=L/2$, i.e.,
half-filling. If the hoppings are homogeneous, $t_i=1$, the chain is called
{\em clean} and can be described in the continuum limit by a conformal
field theory (CFT) \cite{DiFrancesco}. Please notice that we do not
consider on-site disorder, i.e. inhomogeneities in the chemical
potential.

The ground state (GS) of \eqref{eq:ham} can always be written as a
Slater determinant:

\beq
\ket|\Psi>=\prod_{k=1}^{N_e} b^\dagger_k \ket|0>,
\label{eq:slater}
\eeq
with $\ket|0>$ the Fock vacuum and $b^\dagger_k$ the creation
operators for the orbitals, given by a canonical transformation

\beq
b^\dagger_k = \sum_i U_{ki} \ c^\dagger_i,
\label{eq:canonical}
\eeq
where $U$ is the matrix that diagonalizes the hopping matrix, $T_{ij}
= t_i \ (\delta_{i,i+1} + \delta_{i,i-1})$, with eigenvalues
$\varepsilon_k$. The energy gap of the system is given by the minimal
excitation energy:

\beq
\Delta E=\varepsilon_{(L/2)+1}-\varepsilon_{L/2}.
\label{eq:gap}
\eeq

The correlation matrix, defined as

\beq
C_{i,j}=\bra<\Psi| c^\dagger_i c_j \ket|\Psi>.
\label{eq:corr}
\eeq
allows us to compute the expectation value of any observable on any
state given by Eq. \eqref{eq:slater}, via Wick's theorem. It can be
evaluated using the matrix $U$:

\beq
C_{i,j}=\sum_{k=1}^{N_e} \bar U_{ki} U_{kj}.
\label{eq:corr_slater}
\eeq

Notice that the local density, $\<n_i\>=\<c^\dagger_i c_i\>$, is given
by the diagonal elements of $C_{i,j}$. Making use of chiral symmetry
it can be proved that, at half-filling, the density must be
homogeneous, i.e. $\<n_i\>=1/2$ for all $i\in \{1,L\}$ \cite{Asboth}.

\subsection{Entanglement of free fermionic chains}

Entanglement properties of a generic block of the chain,
$B=\{i_1,\cdots,i_\ell\}$ (note that the sites $i_1,\cdots,i_\ell$ are
possibly disjointed), are always referred to the reduced density
matrix of $\ket|\Psi>$, defined as

\beq
\rho^B \equiv 
\Tr_B \ket|\Psi>\bra<\Psi|  ,
\label{eq:rho}
\eeq
being $\Tr_B$ the corresponding partial trace. In the case of a Slater
determinant, this $\rho^B$ can be expressed as a tensor product of
$2\times 2$ density matrices of the form \cite{Peschel.03}

\beq
\rho^B = \bigotimes_{k=1}^\ell
\pmat{\nu^B_k & 0
  \\ \\ 0 & 1-\nu^B_k}.
\eeq
where the $\nu^B_k \in [0,1]$ are the eigenvalues of the
correlation $\ell\times\ell$ sub-matrix corresponding to the block
$C^B$ (i.e., those elements of $C_{i,j}$ with $i, j \in B$).

As a measure of the entanglement between the block and the rest of the
system we choose the von Neumann entropy of $\rho^B$,

\beq
S^B = 
- \Tr \rho^B \log\rho^B,
\label{eq:Srho}
\eeq
which can be computed using the following expression \cite{Peschel.03}

\beq
S^B = -\sum_{k=1}^\ell 
\Big( 
\nu_k^B \log\nu_k^B + 
(1- \nu_k^B) \log(1-\nu_k^B) 
\Big).
\label{eq:S}
\eeq

Our interest in the aforementioned entanglement measures stems from
the fact that they are usually able to characterize the different
phases of matter through e.g. corrections (or violations) to the
area law for the entanglement entropy \cite{Vidal.03,Ramirez.14}.

It is also relevant to ask about the spatial origin of entanglement
within a block. An {\em entanglement contour} $s^B(i)$ \cite{Vidal.14}
is defined as a distribution of the block entanglement among the sites
with some obvious properties, such as positivity and completeness:

\beq
\sum_{i=1}^\ell s^B(i) = S^B, \qquad
s^B(i)\geq 0.
\label{eq:contour_prop}
\eeq
A concrete proposal for an entanglement contour for free fermions was
made by Chen and Vidal \cite{Vidal.14}, which has been shown to match
the CFT prediction \cite{Tonni.18}. The expression is given by

\beq
s^B(i)=-\sum_{k=1}^\ell |V^B_{ki}|^2
\big( 
\nu_k^B \log\nu_k^B + 
(1- \nu_k^B) \log(1-\nu_k^B) 
\big).
\label{eq:contour}
\eeq
where $V^B_{ki}$ is the $i$-th component of the $k$-th eigenvector of
the correlation sub-matrix $C^B$. Both properties in
Eq. \eqref{eq:contour_prop} are straightforward to prove. Let us
remark that the entanglement contour has already been employed to
provide insight into the study of quantum phases of matter
\cite{Tonni.18,Alba.18}.

\subsection{Dasgupta-Ma Renormalization}

When the values of the hoppings are very different among themselves, a
useful approximation is provided by the Dasgupta-Ma renormalization
scheme \cite{Dasgupta.80,Ramirez.14b}, a second-order perturbation
theory approach which was initially devised for random spin chains and,
thus, it is known as {\em strong disorder renormalization group}
(SDRG). We should stress that the main requirement for the
applicability of the SDRG is not disorder, but strong inhomogeneity,
as shown in the applicability to e.g. the rainbow chain
\cite{Ramirez.14,Ramirez.15,Laguna.16}. When the inhomogeneity of the
hoppings is not so strong, the accuracy of the SDRG algorithm will
decrease. Yet, it has been shown in a variety of cases that as the
inhomogeneity is decreased, the exact GS undergoes a smooth crossover
between the SDRG prediction and the homogeneous (or clean) GS
\cite{Ramirez.14b,Ramirez.15,Laguna.17,Tonni.18}.

In order to obtain the GS of Hamiltonian \eqref{eq:ham} on a generic
system with strongly inhomogeneous hoppings, the SDRG proceeds in an
iterative way by always selecting the strongest link and putting a
valence bond between the two sites with this strongest link. Next,
the neighboring sites to this bond are linked by an effective hopping
which is found using second-order perturbation theory
\cite{Dasgupta.80,Ramirez.15}:

\beq
t^\eff = 
- \frac{t_L \ t_R}{t_{\mathrm{\max}} },
\eeq
where $t_L$ and $t_R$ are the left and right hoppings, and $t_\max$ is
the maximal hopping (in absolute value). Notice that the effective
hopping can have different signs. At half-filling, the algorithm
proceeds until all sites are part of one of such valence bonds. Thus,
the GS can be described as a valence bond solid (VBS). The energy gap,
$\Delta E$, can be estimated making use of the SDRG. It corresponds to
the (effective) hopping of the last bond \cite{Ramirez.14b}.

It has been recently proved \cite{Alba.18} that, in the case of
free fermions, when a bond is formed between sites $i$ and $j$ of a 1D
chain, the effective hopping within SDRG is always given by

\beq
t_{i,j}^\eff = 
\frac{t_i\; t_{i+2}\; t_{i+4} \cdots t_{j-1} }%
{t_{i+1} \cdots t_{j-2}},
\label{eq:effcoup}
\eeq
i.e. the product of the odd hoppings divided by the product of the
even ones. Thus, when a valence bond is established between the two
extreme sites of a chain, it will usually have the lowest energy and
Eq. \eqref{eq:effcoup} provides an estimate for the energy gap of the
system.


\section{Dimerized chains}
\label{sec:dimerized}

Let us consider a dimerized version of the Hamiltonian given by
Eq. \eqref{eq:ham}, using

\beq
t_i = 1+(-1)^i \ \delta ,
\label{eq:dimer}
\eeq
with $\delta\in [0,1)$. Thus, the first and last hoppings are always
  weaker, $t_1=t_{L-1}=1-\delta$. This is, indeed, the
  Su-Schrieffer-Heeger (SSH) Hamiltonian
  \cite{Su.79,Heeger.88,Sirker.14} specialized for the topologically
  non-trivial phase \cite{Asboth}, where an edge state appears between
  the first and last sites. See Fig. \ref{fig:illust_1} for an
  illustration. For large enough $\delta$, the fermions minimize their
  energy by localizing on valence bonds on top of the stronger links,
  $t_{2k}=1+\delta$, for all $k<L/2$. After all these bonds are
  formed, the remaining fermion, being unable to be localized on those
  hoppings, will be delocalized around the sites 1 and $L$.
  
This dimerization appears naturally in 1D systems due to the {\em Peierls
  distortion} \cite{Peierls.qts,Peierls.surprises}: the coupling
between electrons and phonons on 1D lattices can give rise to a new
spatial periodicity, twice the original one, opening a gap at the
Fermi energy. For example, a quasi-1D polymer like
\textit{trans}-polyacetylene is electrically insulating whereas the
configuration with all links equivalent is metallic. The Peierls
transition is a widespread phenomenon in quasi 1D systems, making
dimerized materials more energetically favorable than other structural
phases in many occasions \cite{Gruner.88,Snijders.10}.

\begin{figure*}
\begin{center}
\includegraphics[width=6.5cm]{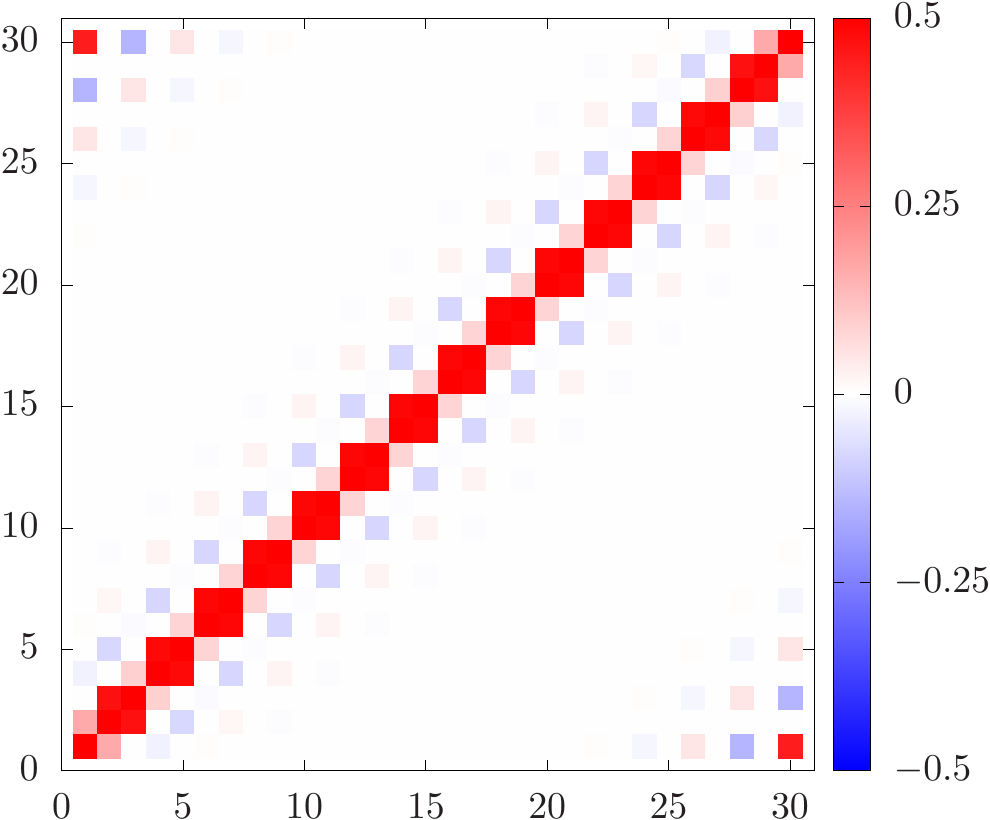}
\includegraphics[width=8cm]{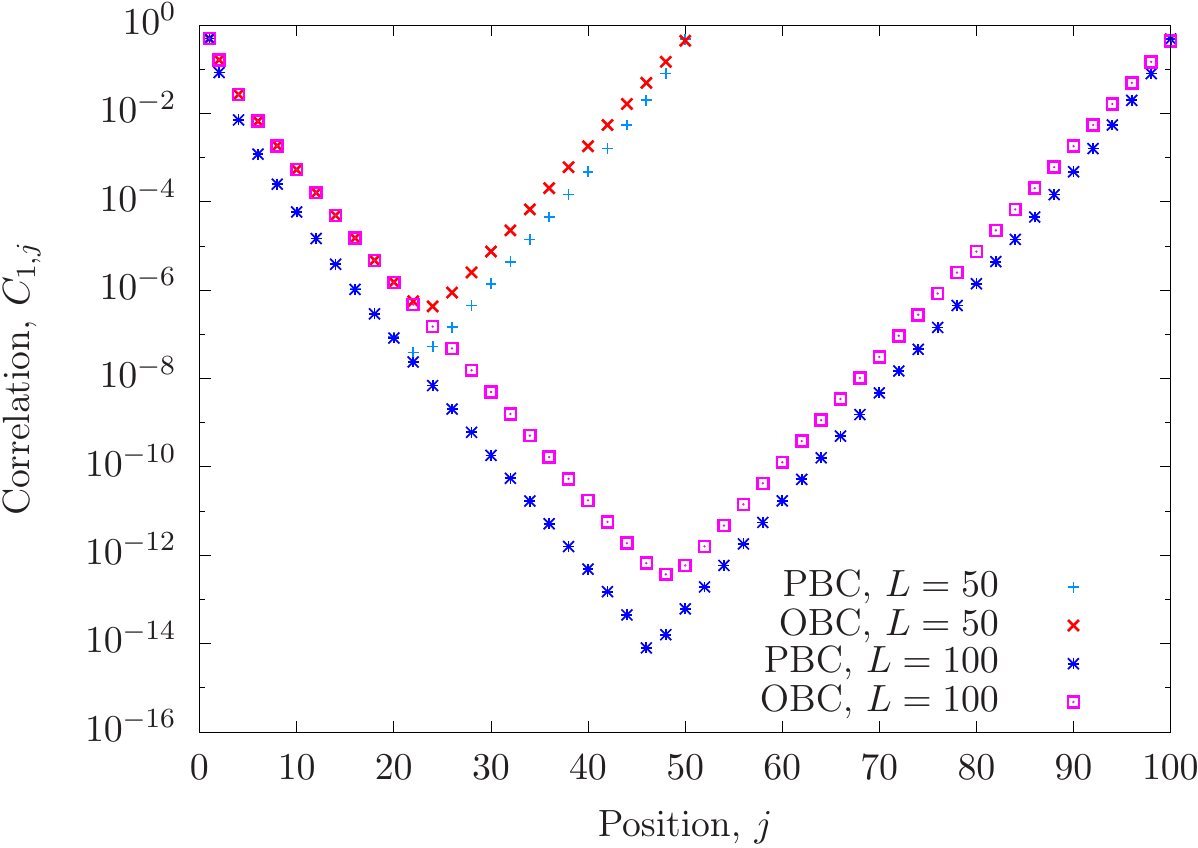}
\end{center}
\begin{center}
\includegraphics[width=7.5cm]{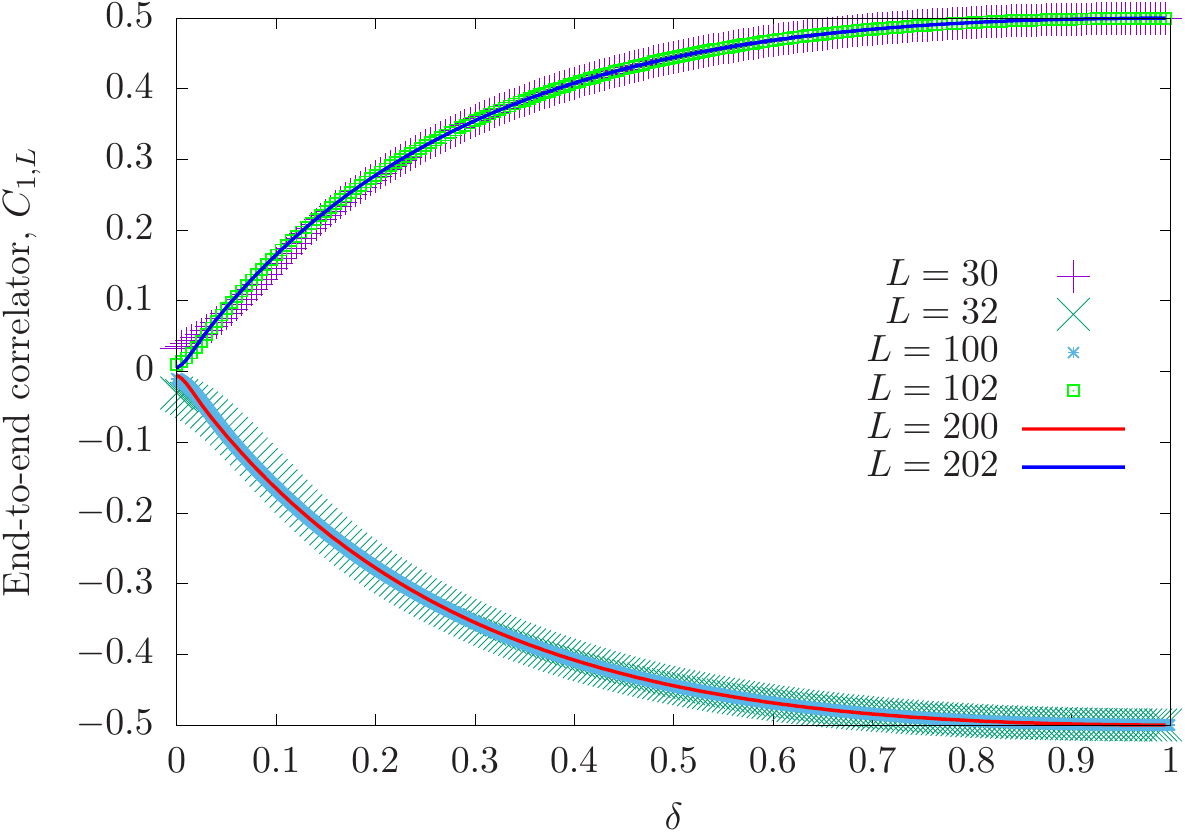}\hbox to 1cm{\hfill}
\includegraphics[width=7.5cm]{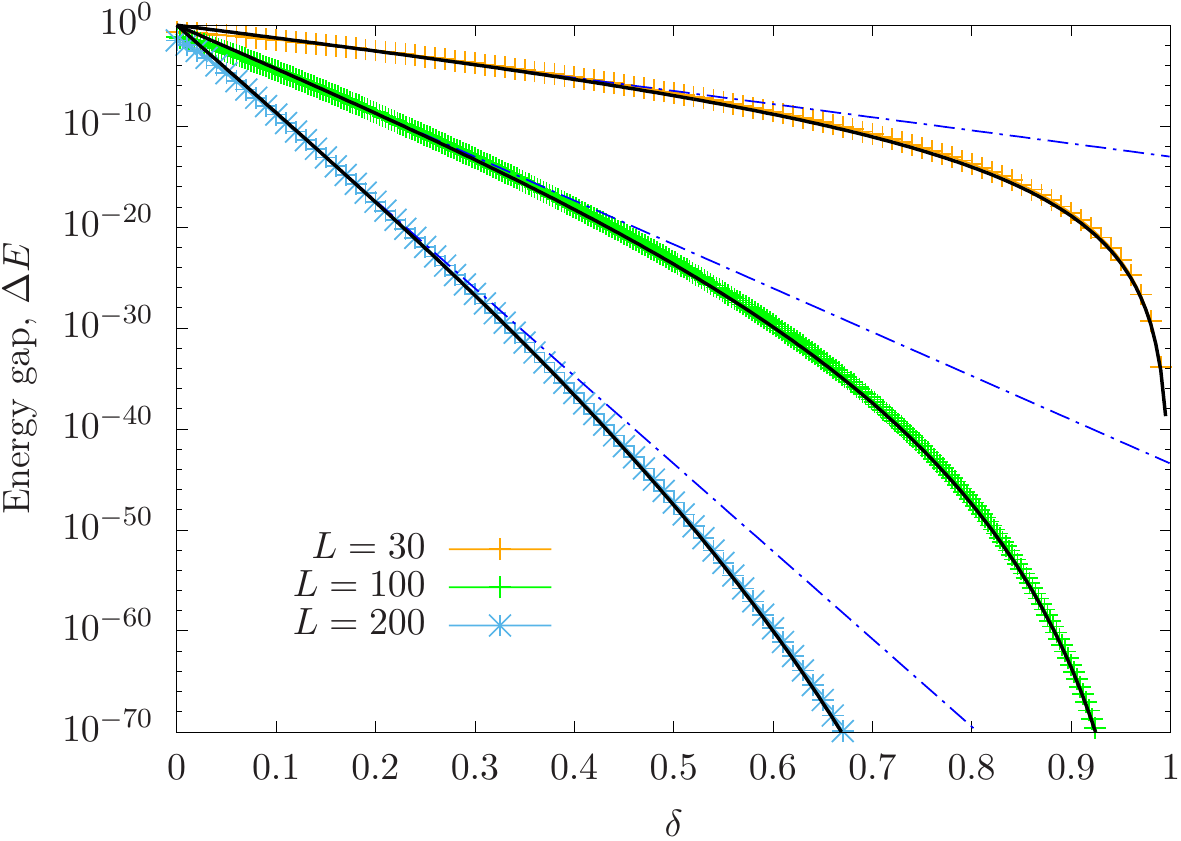}
\end{center}
\caption{Top-left: Correlation matrix, $C_{i,j}$ for the GS of the
  $L=30$ open dimerized chain with hoppings given by
  Eq.~\eqref{eq:dimer} and dimerization parameter $\delta=0.5$. Notice
  the dimerization pattern along the secondary diagonals, and the
  large non-local correlations between sites $1$ and site $L$.
  Top-right: Absolute value of the correlation function $C_{1,j}$ in
  logarithmic scale, for both open (OBC) and periodic (PBC) boundary
  conditions, for $L=50$ and $L=100$ and $\delta=0.5$.  Notice the
  similarity between them.  Bottom-left: end-to-end correlation
  obtained in an open dimerized chain of the form Eq. \eqref{eq:dimer}
  as a function of $\delta$ for several values of $L$, showing the
  collapse.  Bottom-right: energy gap of the system; the dashed
  straight lines represent exponential decays,
  $\exp(-L\ \delta/2)$. The continuous black lines correspond to the
  simple theoretical estimate given by Eq.~\eqref{eq:dimergap}.  }
\label{fig:dimer}
\end{figure*}

The top-left panel of Fig.~\ref{fig:dimer} presents the correlation
matrix $C_{ij}$ of a chain with $L=30$ sites when $\delta=0.5$,
as computed using Eq.~\eqref{eq:corr_slater}. Notice that all the
diagonal elements equal $1/2$, because in this case the fermionic
density at the sites can be proved to be homogeneous (summing up all
occupied states, we have a constant local density of states on all the
sites of the system, even with this distinct correlation pattern). The
off-diagonal elements, then, show the correlations in our system. All
matrix elements of the form $C_{2k,2k+1}$ take a much higher value
than those of the form $C_{2k-1,2k}$, showing a strong
dimerization. The value $C_{1,L}$ is also high, signaling the expected
presence of an edge-state \cite{Asboth}. We will call that term the
{\em end-to-end correlation}.

The top-right panel of Fig.~\ref{fig:dimer} shows, in logarithmic
scale, the absolute value of the correlation between site 1 and all
others, comparing two sizes ($L=50$ and $100$) and two types of
boundary conditions (open (OBC) and periodic (PBC)).  Note that the
results for the correlation function $C_{1,j}$ at the end of the
chains (i.e., for $j=50$ and $100$, respectively) are about $1/2$ for
both open and periodic conditions. The periodic case is well known:
$C_{1,j}$ falls exponentially until $j \sim L/2$ (the center of the
chain), where it becomes quite small; then it raises exponentially
again. Interestingly, the same large decay and increase takes also
place for open boundaries.

In the bottom-left panel of Fig.~\ref{fig:dimer} we have plotted this
end-to-end correlation, $C_{1,L}$ as a function of the dimerization
parameter $\delta$ for open systems of different sizes. The
correlation sign depends on whether the value of $L \inmod 4$ equals 0
or 2, because of the parity of the number of fermions in the chain
bulk.  The collapse of all curves with the same parity for different
system sizes is remarkable.

It is relevant to ask about the stability of these large end-to-end
correlations, i.e., what is the effective hopping associated to them,
directly related with the energy gap $\Delta E$, given by
Eq. \eqref{eq:gap}. The bottom-right panel of Fig.~\ref{fig:dimer}
shows this energy gap as a function of $\delta$ for different system
sizes.  Notice the exponential decay for small $\delta$, becoming even
faster for larger dimerizations. The dashed lines correspond to the
exponential behavior,

\beq
\Delta E\sim \exp(-L\,\delta/2).
\label{eq:gapdimer}
\eeq
Yet, the behavior of the energy gap along all the range for $\delta$
can be successfully estimated making use of the SDRG
approximation. Indeed, Eq. \eqref{eq:effcoup} can be applied to our
case, using the expression for the hoppings given in
Eq. \eqref{eq:dimer}, and obtaining

\beq
t_{1,L}^\eff = 
\frac{ (1-\delta)^{L/2} }{ (1+\delta)^{L/2-1} }.
\label{eq:dimergap}
\eeq
This effective SDRG hopping between the two extremes is a
good approximation for the energy gap as shown by the black continuous
curves on the bottom-right panel of Fig. \ref{fig:dimer}, which fit
the numerical results for $\Delta E$ with remarkable accuracy over
several orders of magnitude.

Unfortunately, this result also leads to a predictable conclusion:
since the energy gap $\Delta E$ is so small, the edge states of the
SSH model are extremely fragile.

\subsection{Entanglement properties of the dimerized chain}

The introduction of dimerized hoppings on a clean free-fermionic
infinite chain decrease notably the correlations on the GS, see the
bottom-left panel of Fig. \ref{fig:dimer}. Moreover, the appearance of
an energy gap forces the state to fulfill the area-law
\cite{Hastings}, thus making the maximal entropy bounded. It is
interesting to consider the entanglement properties of finite SSH
chains in order to determine with more accuracy the different
contributions of the bulk and the edge. Remarkably, a low degree of
dimerization has been shown to increase the entanglement between the
left and right halves of the system with respect to the clean case
\cite{Sirker.14,Sedlmayr.18}, and this increase in entanglement
entropy has been suggested as a mechanism behind the Peierls
transition \cite{Sirker.08,Sirker.14}. In this section we intend to
extend previous findings about the entanglement behavior of the SSH
chain \cite{Sirker.14,Sedlmayr.18} with the use of the entanglement
contour \cite{Vidal.14}. This, in turn, will help us in our main aim
of establishing stable quantum chains with large end-to-end
correlations.

\begin{figure}
\includegraphics[width=7.5cm]{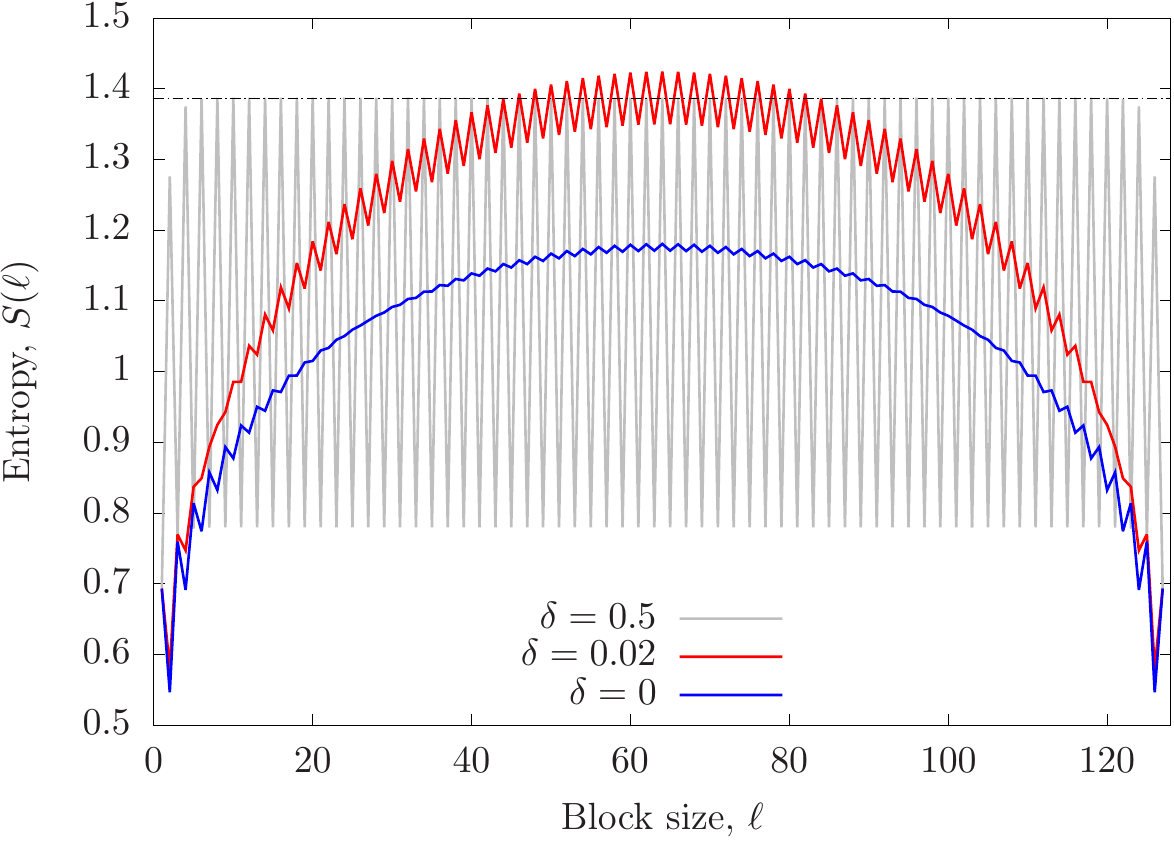}
\includegraphics[width=7.5cm]{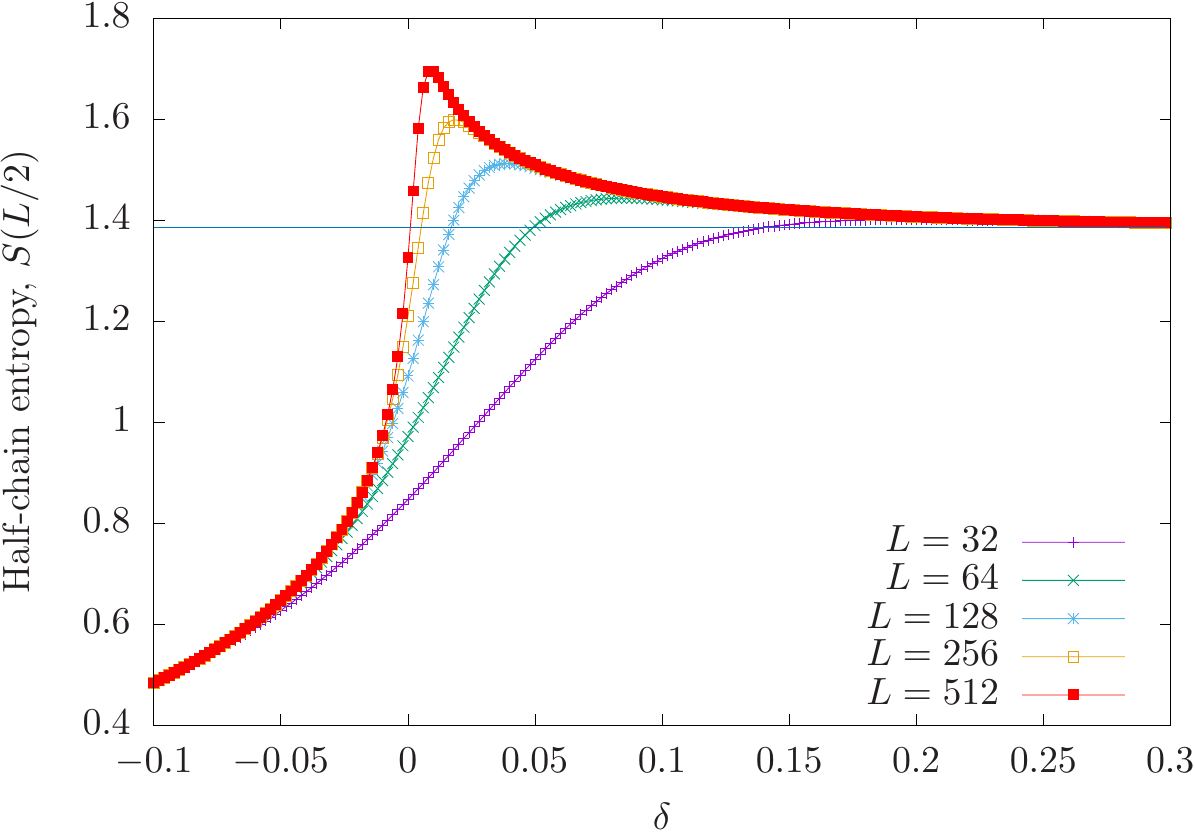}
\includegraphics[width=7.5cm]{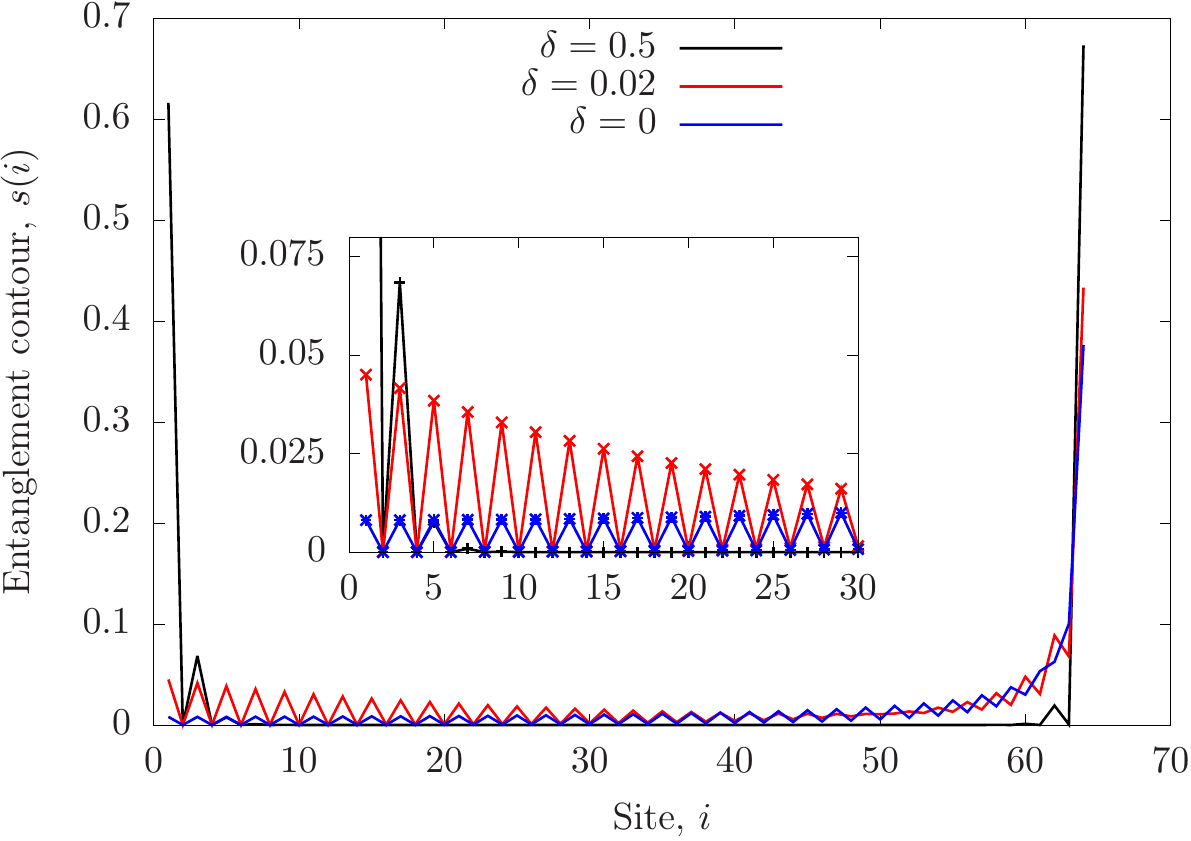}
\caption{Top: Entanglement entropy of blocks $\{1,\cdots,\ell\}$ as a
  function of the block size $\ell$ for different values of $\delta$,
  using $L=128$, on the GS of an open dimerized chain given by
  Eq.~\eqref{eq:dimer}. Center: Half-system entanglement entropy
  $S(L/2)$ as a function of the dimerization parameter, for different
  values of $L$. Notice the initial linear surge, followed by a
  monotonous (exponential) decay towards $2\log(2)$, indicated by
  horizontal lines in both panels. Bottom: entanglement contour for
  the left half of the $L=128$ system and the same values of $\delta$
  used for the top panel. Inset: zoom on the left part of the block.}
\label{fig:entropy}
\end{figure}

The top panel of Fig.~\ref{fig:entropy} shows the {\em entanglement
  entropy} of blocks comprising the $\ell$ leftmost sites, $S(\ell)$,
obtained with Eq.~\eqref{eq:S} for different values of $\delta$ in an
open dimerized chain with $L=128$. In the figure, it can be seen how
the entanglement entropy $S(\ell)$ increases as $\delta$ increases.
The amplitude of the oscillations also increases, as for $\delta\to
1^-$ the system becomes a valence bond solid. For zero dimerization,
$\delta=0$, $S(\ell)$ almost reproduces the well-known form obtained
from CFT \cite{Vidal.03,Calabrese.04},

\beq
S = \frac{1}{6} 
\log \Bigg( \frac{L}{\pi} \sin\( \frac{\pi\ell}{L} \) \Bigg) 
+ Y(\ell)
\label{eq:Scft}
\eeq
where the first term is provided by the CFT and $Y(\ell)$ is a
non-universal mild oscillatory term \cite{Calabrese.09,Xavier.11}.

On the other side, as commented, in the $\delta\to 1^-$ limit the GS
becomes a valence bond solid (VBS). In that regime, entanglement
entropies of given blocks are easy to estimate: one must simply count
the number of broken bonds when the block is separated from the
environment, and multiply by $\log(2)$. In this strong dimerization
regime, the block entanglement entropy $S(\ell)$ becomes exactly
oscillating, alternating values of $\log(2)$ (single bond cut) and
$2\log(2)$ (two bonds cut), as depicted in the physical picture shown
in Fig.~\ref{fig:illust_1}.

The central panel of Fig.~\ref{fig:entropy} shows $S(L/2)$ (i.e. the
entanglement entropies for the half chain) as a function of $\delta$
for several system sizes, always choosing multiples of 4 in order to
have two bond cuts in the strongly dimerized limit. We observe a fast
linear increase of the entropy for low values of $\delta$, reaching a
maximum that increases with the system size. Beyond that maximum, the
entropy for all different sizes collapse to a single curve which
approches asymptotically the limit value $S(L/2) \to 2\log(2)$ for
$\delta \gtrsim 0.15$. Notice also that we have included the
computation for $\delta<0$, obtaining lower values of the entanglement
entropy that also collapse.

It is enlightening to consider the {\em entanglement contour}, defined
in Eq. \eqref{eq:contour}, for the left half of the chain with $L=128$
sites, using $\delta=0$, $0.02$ and $0.5$, as shown in the bottom
panel of Fig. \ref{fig:entropy}. Notice that the left extreme
corresponds with the physical left boundary of the chain, while the
right extreme of the plot corresponds with the center of the chain.
We can see that for large $\delta$ (black curve) the entanglement
contour is large at both extremes, and small everywhere else. The
reason is that only sites $1$ and $L/2$ contribute to the block
entanglement, because they take part in broken valence bonds. For
$\delta=0$ we recover the conformal case \cite{Tonni.18}, where it is
known that the entanglement contour falls as a power law, $s(i)\sim
(L/2-i)^{-1}$. For the selected intermediate value, $\delta=0.02$, we
see that the contour on the right extreme of the block is very similar
to the clean case ($\delta=0$). Nonetheless, we also observe that the
contour on the left half has risen considerably. Thus, we can argue
that the entanglement excess is produced in the bulk, but due to a
boundary effect \cite{Sedlmayr.18}.

Our preliminary conclusions from this study are that (a) dimerization
at the edge of the chain can give rise to strong end-to-end
correlations; (b) the energy gap is enormously reduced due to the bulk
dimerization (see Eq. \eqref{eq:dimergap}) and (c) the initial surge
in entropy when we dimerize a clean chain (see central panel of
Fig. \ref{fig:entropy}) is a bulk phenomenon. These conclusions will
help us in the following sections.


\section{Correlation engineering in chains}
\label{sec:engineering}

Can we obtain large end-to-end correlations in a fermionic chain with
a large enough energy gap?

Let us consider an open chain with $L$ sites and $N_e=L/2$ fermions,
described by Eq.~\eqref{eq:ham}. The absolute value of the end-to-end
correlation, $|C_{1,L}|$ can be regarded as a function of the hopping
amplitudes, $\{t_i\}_{i=1}^{L-1}$. We can now obtain the maximum of
this function with the energy gap constrained to take a fixed value
$\Delta E$, Eq. \eqref{eq:gap}, using an optimization algorithm. In
order to avoid a trivial increase of the energy gap through a
rescaling of the hopping terms, we will normalize it with the average
value of the hoppings, $\bar t$ (all hoppings will be restricted to be
positive).

The aforementioned optimization is a non-trivial task, due to the
complex landscape exhibited by the target function
\cite{Weise,Schrijver}. It has been recently shown that
machine-learning techniques can be suitable for the solution of
quantum many-body problems \cite{Carleo.17,Fujita.18}. In this work we
employ numerical techniques inspired in machine-learning in order to
obtain the desired optimal hoppings in an efficient way.

\subsection{Machine-learning technique}

The optimizer algorithm receives two parameters: the system length,
$L$, and the expected value of the energy gap $\Delta E_0$. The
target function is defined as

\beq
    {\cal F}\(\{t_i\}_{i=1}^{L-1}\) \equiv
    |C_{1,L}|-K\left| {\Delta E - \Delta E_0 \over \bar t} \right|.
\label{eq:target}
\eeq
where $K$ is a {\em constraint coefficient}, which is varied along the
algorithm, ultimately reaching a very large value in order to ensure
that the energy constraint is fulfilled. Notice that the energy gap is
always measured in units of the average hopping.

The algorithm starts out with $N_c$ random configurations (typically,
$N_c=8$) for the hoppings with fixed average, $\bar t=1$. A conjugated
gradients \cite{NRC} search is performed on each of these initial
configurations in three stages, increasing progressively the value of
$K$ (typically, $K_1=1$, $K_2=100$ and $K_3=10^4$). After this
procedure, the best $N_c/2$ configurations are stored with a small
random perturbation. We add $N_c/2$ new random configurations and the
cycle repeats again. After a few cycles (always less than $10$), the
best configuration is selected.

\subsection{Results: optimal chains}

\begin{figure*}
\includegraphics[width=8.5cm]{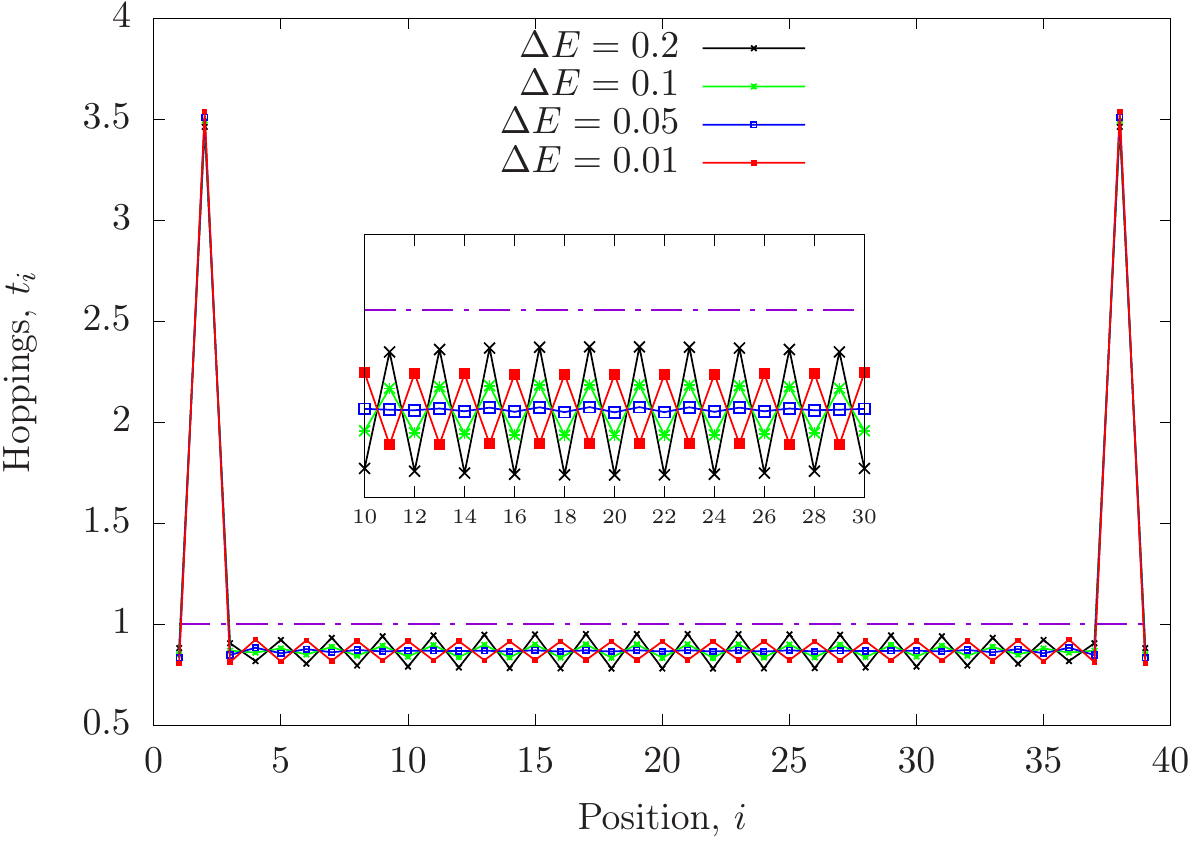}
\qquad
\includegraphics[width=8.5cm]{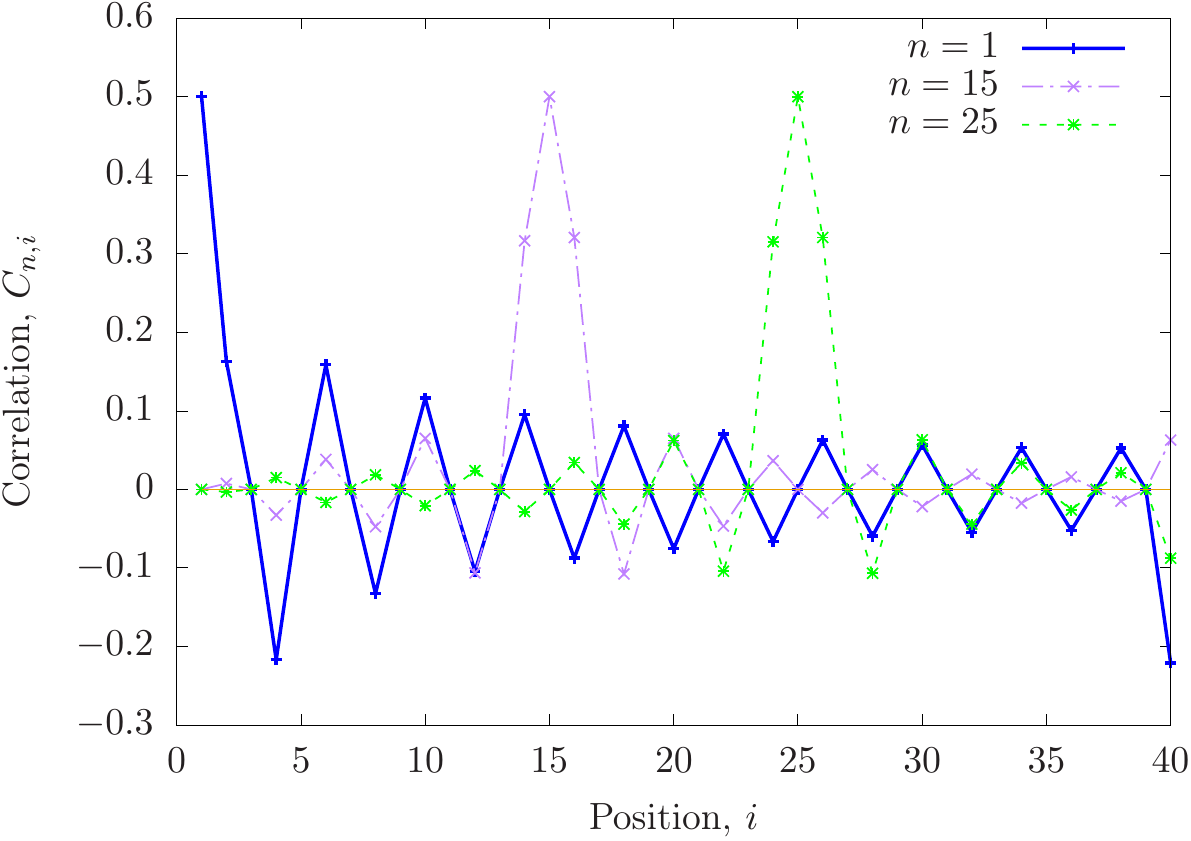}
\vspace{0.5cm}
\includegraphics[width=8.5cm]{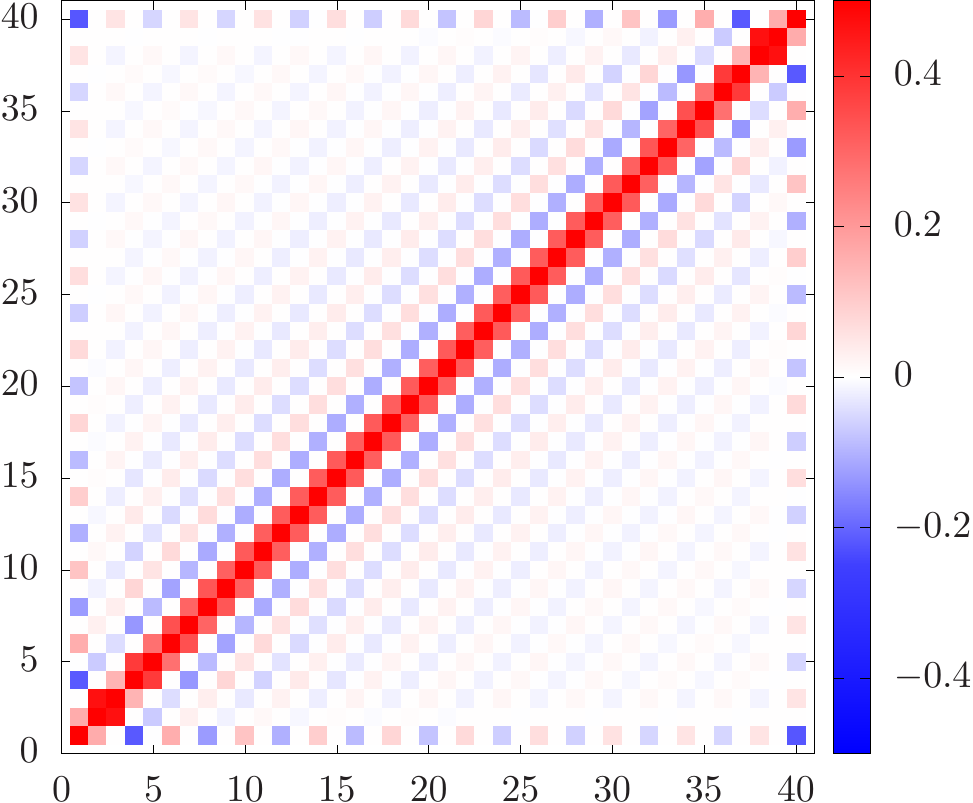}
\includegraphics[width=8cm]{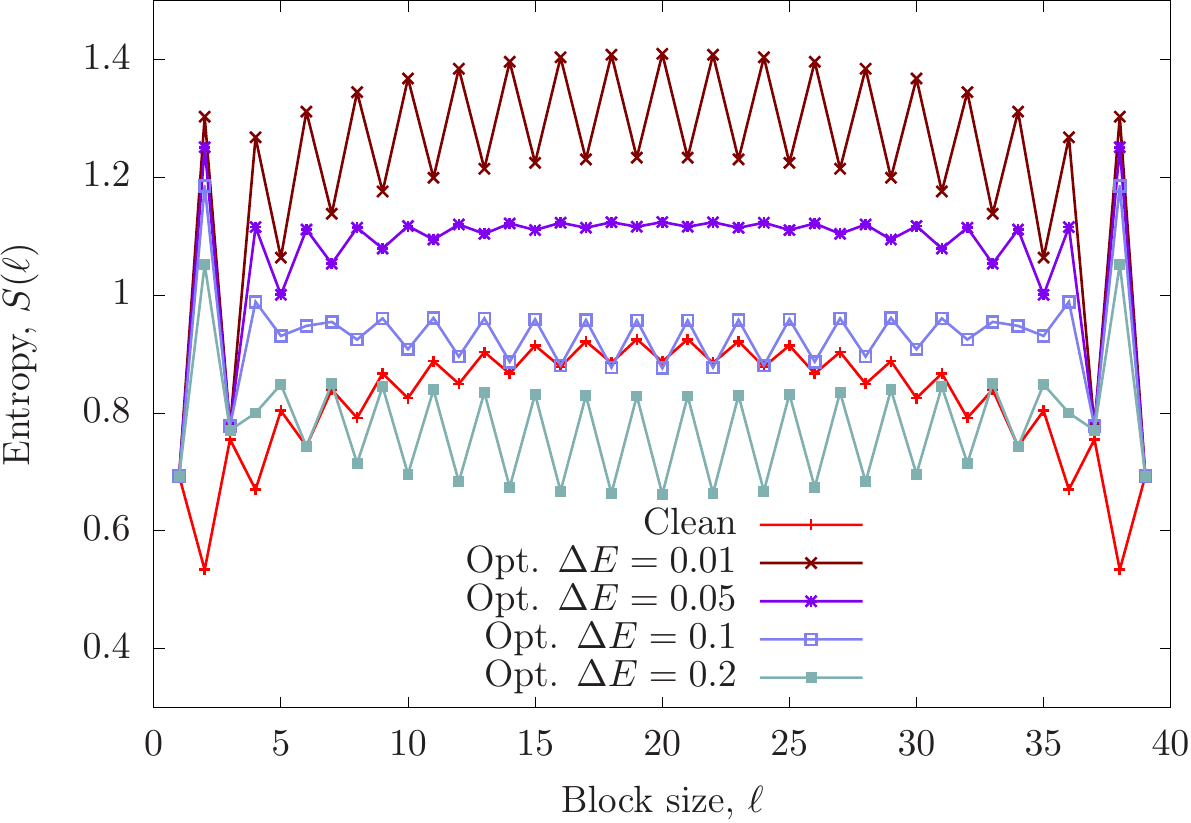}
\caption{Optimizing the end-to-end correlation in an open fermionic
  chain with $L=40$ sites for a fixed gap and average hopping fixed to
  1. Top-left: optimal hoppings for fixed gaps $\Delta E=0.2$, 0.1,
  0.05 and 0.01. Notice that, in all cases, the dimerization is
  strongly modulated, with the second and penultimate hoppings always
  significantly stronger than the rest. Inset: zoom of the previous
  data in the central zone. Notice that the dimerization is {\em
    opposite} in very low and large gaps. Top-right: correlation
  functions $C_{n,i}$ for selected sites, $n=1$, $15$ and $25$ for the
  case $\Delta E=0.05$. The $n=1$ shows a strong peak at the end, with
  an end-to-end correlation $C_{1,L}\approx -0.221$. Bottom-left: full
  correlation matrix for the $\Delta E=0.05$ case. Bottom-right: block
  entropies for the cases shown in the top-left panel.}
\label{fig:engin}
\end{figure*}

In the top-left panel of Fig. \ref{fig:engin} we present the optimal
hopping amplitudes for $L=40$ with a few values of the fixed energy
gap $\Delta E=0.01$, $0.05$, $0.1$ and $0.2$ (always in units of the
average hopping, $\bar t$), which give rise to maximal (negative)
end-to-end correlations $-0.384$, $-0.221$, $-0.107$ and $-0.021$
respectively.  The systematic study of the maximal correlation
achievable for a given gap is performed in the next section. At this
moment, let us explore the resulting hopping profiles.

The optimal hopping profiles shown in the top-left panel of
Fig. \ref{fig:engin} show a strongly modulated dimerization. Indeed,
the second and penultimate links become significatively stronger, with
$t_2=t_{L-2}\approx 3.5$. This {\em edge-dimerization} is present in
the optimal hopping patterns for all target values of $\Delta E$. On
the other hand, the dimerization amplitude is much smaller in the
bulk. The inset of the top-left panel of Fig. \ref{fig:engin} still
shows another interesting surprise: the bulk dimerization takes a
different phase for large and smaller values of the energy gap $\Delta
E$. Indeed, for small gap ($\Delta E=0.01$, red line) the dimerization
phase is the same as in the previous section (pattern
$(1-\delta),(1+\delta),\cdots,(1-\delta)$). But the pattern is
reversed for larger values of the gap. The reason can be understood
via Eq. \eqref{eq:effcoup}. The energy gap grows by shifting the large
hopping values to the numerator and the small ones to the denominator,
at the expense of reducing the end-to-end correlation.

The top-right panel of Fig.~\ref{fig:engin} shows the correlation
function, $C_{n,i}$ for several values of $n$ ($n=1$, $15$ and $25$)
in the case of $\Delta E=0.05$, when the dimerization amplitude is
minimal among all the cases shown in the left panel. We observe that
the end-to-end correlation is negative, $C_{1,L}=-0.221$, and that the
amplitudes of the correlation function, $C_{1,i}$, decay very slowly
with the distance from the origin. For the other cases, $C_{15,i}$ and
$C_{25,i}$, we see a much faster decay. The bottom-left panel of
Fig. \ref{fig:engin} presents the correlation data for the previous
case ($L=40$, $\Delta E=0.05$) in matrix form, where we can observe
the large end-to-end correlation in the upper-left and lower-right
corners.

Therefore, we can conjecture the following theoretical propostion: if
some sites form strong bonds with their neighbors and as a consequence
some sites are left isolated, these isolated sites are {\em forced} to
establish large correlations between them, even if they are at a long
distance. Of course, these long-distance bonds will be weaker. As
discussed above, the energy gap can be estimated using the effective
hopping amplitude obtained through the generalized SDRG expression,
Eq.~\eqref{eq:effcoup}. This effective hopping amplitude takes a large
value when only the second and penultimate hoppings are large, while
the rest are all equal. This result leads to the conjecture that {\em
  edge-dimerized} chains will be always close to providing robust
optimal correlations.

The bottom-right panel of Fig. \ref{fig:engin} shows the block
entanglement entropies of the optimal configurations obtained,
$S(\ell)$, compared to the homogeneous case given by CFT,
Eq. \eqref{eq:Scft}. We can observe that for very low $\Delta E=0.01$
the entanglement shifts vertically and acquires stronger parity
oscillations. The vertical shift ($\sim \log(2)$) is due to the long
bond between the two extremes of the chain, and the parity effect due
to the dimerization. When $\Delta E=0.05$ the hoppings on the bulk
presented the minimal dimerization, and we can see a corresponding
flattening of the entropy curve. In all cases we can see that the
entropy of the block of 2 sites, $S(2)\sim 2\log(2)$. This means that
both site 1 and 2 establish a valence bond with some other sites. In
fact, site 1 attempts to establish the bond with site $L$, while site
2 is strongly connected to site 3. This fact is checked by observing
that the entropy of the block of 3 sites, $S(3)\sim \log(2)$, because
this block contains now a full valence bond plus the broken bond
corresponding to site 1. For larger values of $\Delta E$, we see the
entropy decaying while keeping the edge behavior.

\subsection{Scaling limit: larger optimal chains}

The previously discussed results have an intrinsic interest, since
they allow us to engineer robust devices of nanoscopic scale ($\sim
50$ atoms) with large correlations. Yet, it is relevant to ask whether
this effect can be extended to larger system sizes.

\begin{figure}
  \includegraphics[width=8cm]{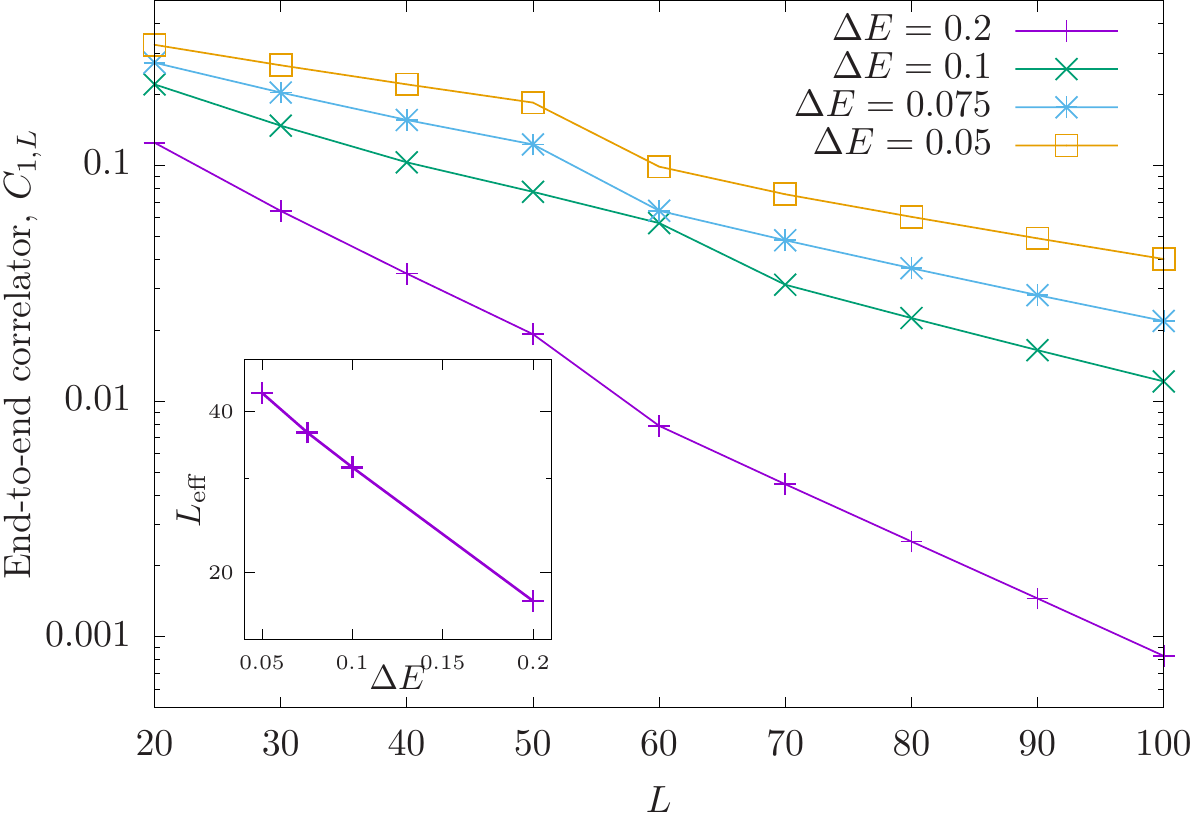}
  \caption{Scaling behavior of the optimal correlation $C_{1,L}$,
    obtained as a function of the system size $L$, for given values
    of the (fixed) energy gap $\Delta E$. In all cases, the maximal
    correlation decays exponentially with the system size,
    Eq. \eqref{eq:corrgap}. Inset: behavior of the effective length
    with the energy gap, $L_\eff(\Delta E)$.}
  \label{fig:tdlimit}
\end{figure}

Fig. \ref{fig:tdlimit} shows the maximal attainable correlation as a
function of the system size, $L$, for different values of the energy
gap, $\Delta E$. In all cases, the maximal correlation decays
exponentially, with a certain effective length which depends on the
energy gap,

\beq
|C_{1,L}| \sim \exp\(-L/L_\eff(\Delta E)\).
\label{eq:corrgap}
\eeq

It is remarkable that, despite the general exponential trend of all
curves in Fig. \ref{fig:tdlimit}, they all present some {\em glitch}
at a finite value of $L$. This is typically due to a change in the
dimerization pattern of the optimal configuration. The dependence of
the effective length on the energy gap, $L_\eff(\Delta E)$, is shown
in the inset of Fig. \ref{fig:tdlimit}. We can see that it also decays
exponentially:

\beq
L_\eff \sim \exp\(-\Delta E/\Delta E_*\),
\eeq
where $\Delta E_* \sim 0.17$.

Therefore, the results presented in this section lead us to conjecture
that, in order to obtain the maximal end-to-end correlation in a
fermionic chain keeping a large energy gap, the best strategy is
usually to induce a strong dimerization at the edge with a homogeneous
bulk. Yet, the maximal correlation for a fixed energy gap will always
decrease exponentially.


\section{Modulated and Edge Dimerization}
\label{sec:modulated}

\subsection{Modulated dimerized chains}
\label{sub:modulated}

The results regarding the optimized correlations presented in
Section~\ref{sec:engineering} hint at a conjecture: a {\em modulated}
dimerization may achieve strong end-to-end correlations with a broader
energy gap. We will explore that conjecture along this section. The
modulation is achieved by allowing the dimerization parameter to vary
along the open boundary chain. Let us introduce a continuous
modulation function $\delta(x): [-1,1]\mapsto \R^+$ and its discretized
version: 

\beq
\delta_i = \delta\( {i-L/2 \over L/2} \).
\label{eq:deltafunction}
\eeq
Now, let the Hamiltonian take the following form:

\beq H = -\sum_i^{L-1} \big( 1 + (-1)^i \ \delta_i \big) \; c^\dagger_i c_{i+1},
\label{eq:modulate}
\eeq
Notice that, by construction, the average value of the hopping terms
is always one. Thus, the energy scale is fixed, and we can use the
energy gap in the spectrum to measure the stability of the GS. Also
notice that, in all cases, the first and last hoppings will be weaker.

We have explored several possibilities for the dimerization function
$\delta(x)$, always increasing the dimerization towards the extremes,
and symmetrical with respect to the center of the chain.
\vspace{2mm}

{\bf (a)} No modulation, as in Sec. \ref{sec:dimerized},
  $\delta(x)=\delta_0$.
\vspace{2mm}

{\bf (b)} Linear modulation, $\delta(x) = \delta_0 \ |x|$.
\vspace{2mm}

{\bf (c)} Quadratic modulation, $\delta(x) = \delta_0 \ x^2$.
\vspace{2mm}

{\bf (d)} Exponential modulation, given by the expression

\beq
\delta(x)= \delta_0 \exp\(\lambda (|x|-1)\).
\label{eq:expmod}
\eeq

\vspace{2mm}

\begin{figure}
  \includegraphics[width=8.5cm]{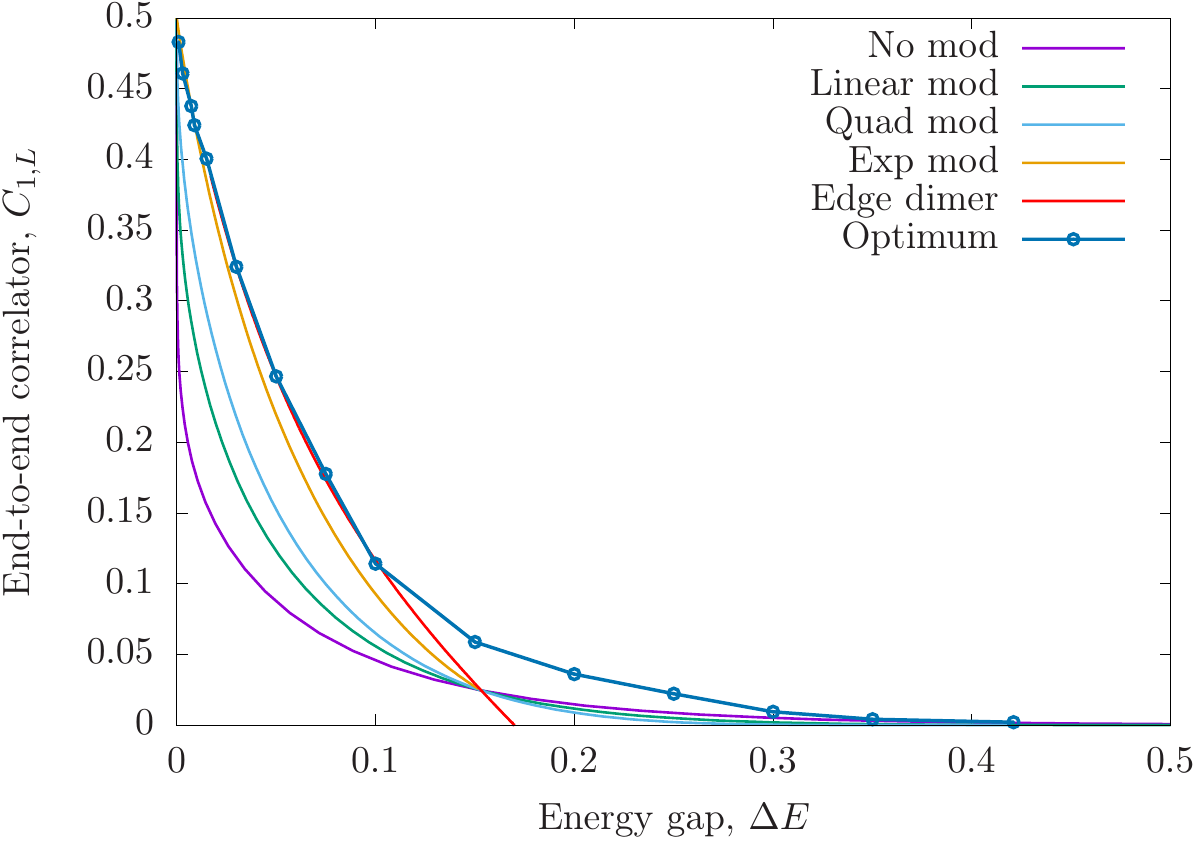}
  \caption{End-to-end correlation in open chains, $C_{1,L}$, as a
    function of the energy gap, $\Delta E$, for different chain types
    whose hoppings are subject to a modulated dimerization,
    Eq.~\eqref{eq:modulate}. Different values of $\delta_0$ (or $t_0$)
    lead to different points of the curve. The different modulation
    types are: no-modulation; linear modulation; quadratic modulation;
    exponential modulation (Eq. \eqref{eq:expmod} with
    $\lambda=8$). All these cases are compared to the {\em optimum},
    obtained using the machine learning procedure of
    Sec.~\ref{sec:engineering} (thicker line).  We also show the
    edge-dimerization results discussed in Sec.~\ref{sub:edge} (red
    line) that present very similar results to the optimal hoppings
    for energy gaps $\Delta E \lesssim 0.125$.}
  \label{fig:modulated}
\end{figure}

In Fig. \ref{fig:modulated} we present the results for the
relationship between the energy gap, $\Delta E$, and the end-to-end
correlation, $C_{1,L}$, for these different dimerization schemes on
open fermionic chains using the Hamiltonian \eqref{eq:modulate}

We can see that, in all cases under study, the end-to-end correlation
decreases with the gap. We have added a last curve, given by the {\em
  optimal correlation} for each given value of the energy gap, as
obtained through the machine-learning algorithm of
Sec. \ref{sec:engineering}. We should remark that the optimal curve
remains above the correlation curve for all types of modulation, as
expected. Yet, we can see that the exponential modulation (yellow
line) is, among the proposed functions $\delta(x)$, the one that comes
closest to the optimal one. Again, we see that strong dimerization
near the edges and weak dimerization in the bulk leads to larger
values of the end-to-end correlation, for a fixed energy gap. Thus, it
is natural to take the next step, and dimerize only the {\em edges} of
the chain.

\subsection{Edge-dimerized chains}
\label{sub:edge}

Let us consider an open fermionic chain of $L$ (even) sites, where all
hoppings $t_i=1$ except two, $t_p=t_{L-p}=t_0$.  We will only use
small values of $p$ (1 to 6) in order to study the effect of the
dimerization process near the edges. The average value of the hopping
is $1+t_0/(N-2)$, which is slightly greater than 1. Nonetheless, the
excess energy becomes negligible for enough large sizes.

Firstly, we have obtained the gap and end-to-end correlation for $p=2$
and different values of $t_0$. The results are presented as an added
curve in Fig.~\ref{fig:modulated}, labeled as {\em edge
  dimer}. Different values of $t_0$ lead to different points of the
curve. We can see that for a wide range of gap values, up to $\Delta E
= 0.125$, this curve is very close to the optimal one.

\begin{figure}
\includegraphics[width=8.5cm]{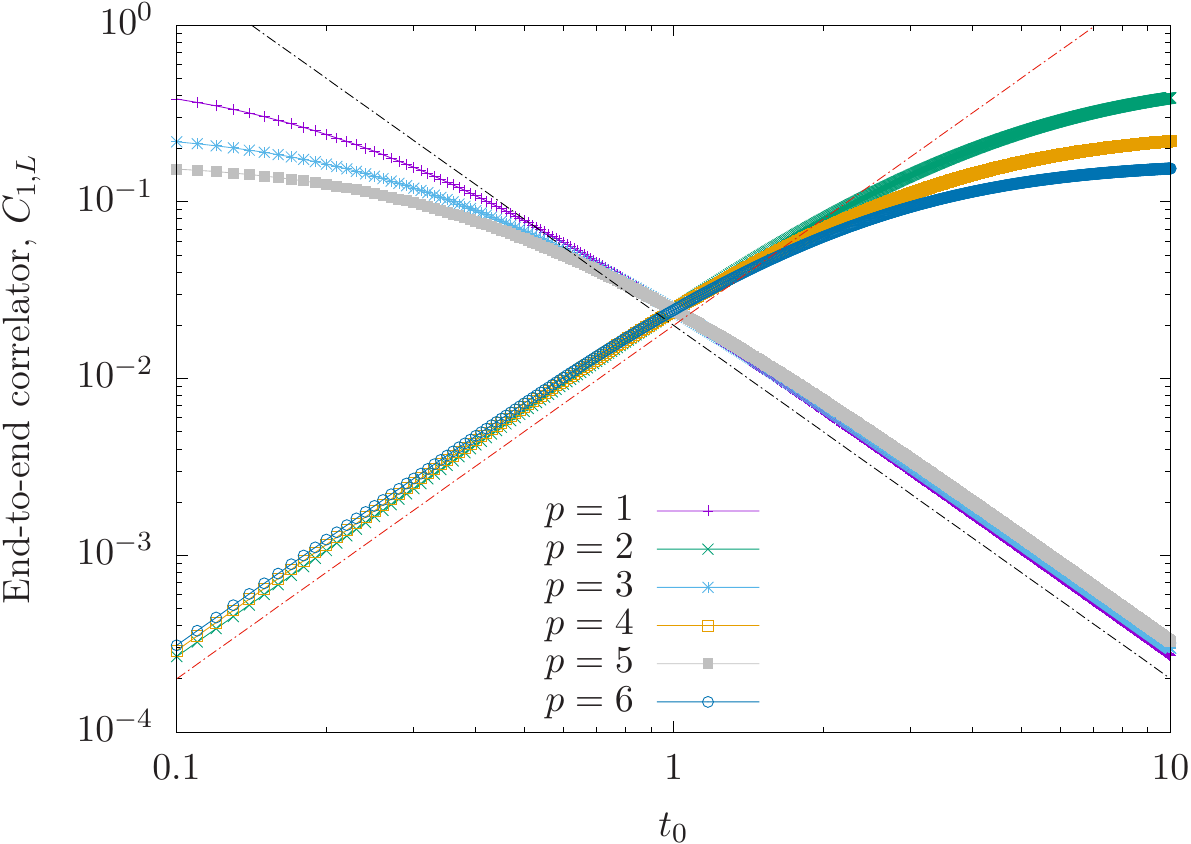}
\includegraphics[width=8.5cm]{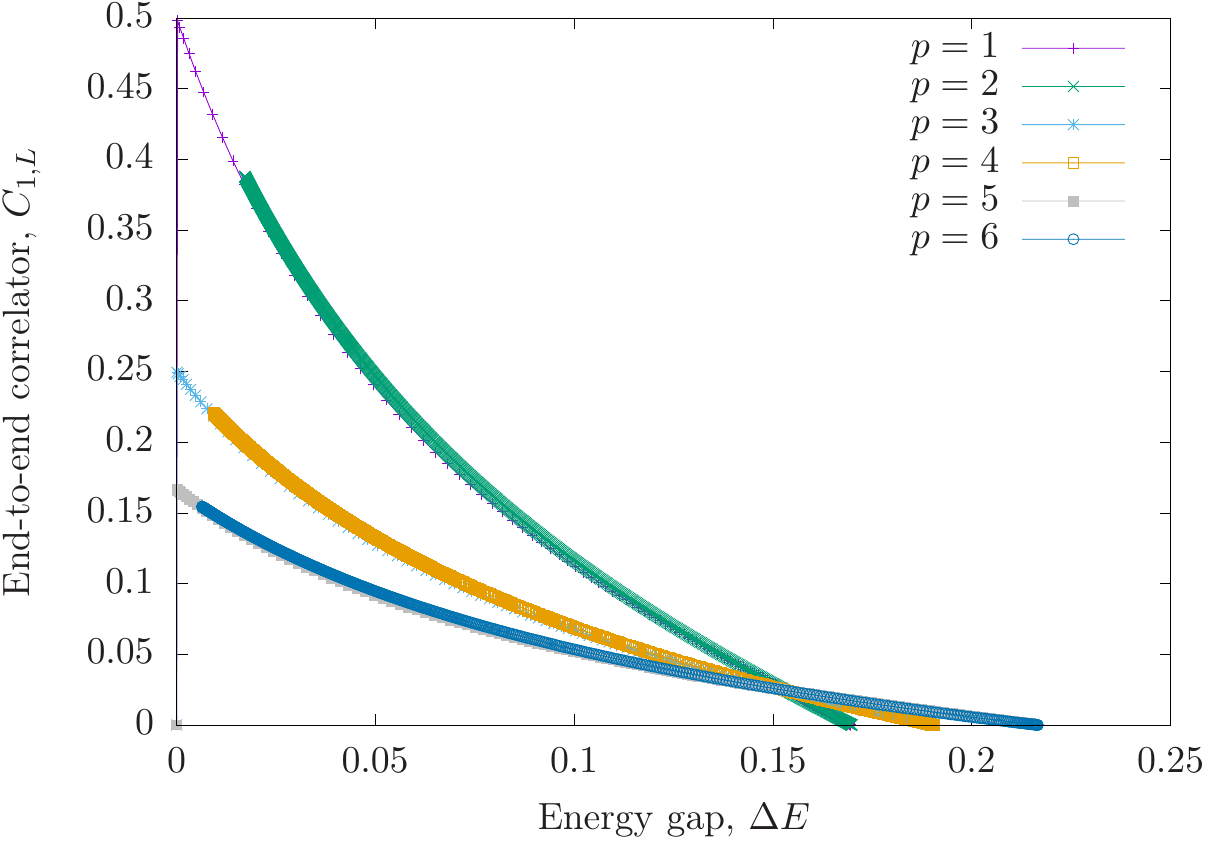}
\caption{Edge-dimerized open boundary chains. Top: End-to-end
  correlation, $|C_{1,L}|$ as a function of the marked hopping, $t_0$,
  for $p=1$ to $6$, in logarithmic scale using $L=40$. The straight
  lines are power-laws corresponding to $t_0^2$ and
  $t_0^{-2}$. Bottom: End-to-end correlation, $|C_{1,L}|$ vs energy
  gap $\Delta E$ in all the previous cases. Notice the collapse of
  all the cases $p=2m-1$ and $p=2m$.}
\label{fig:ed}
\end{figure}

In Fig.~\ref{fig:ed} we show some other properties of the
edge-dimerized open chains. In the top panel we show the end-to-end
correlation as a function of $t_0$ for $p$ from 1 to 6, in logarithmic
scale. Notice that the parity of $p$ is crucial: if $p$ is odd, then
the correlation is strong for low $t_0$ and weak for high $t_0$. The
opposite is true for even $p$.  Moreover, in the low correlation end,
the behavior is a power-law: $C_{1,L}\sim t_0^2$ for even $p$ and
$C_{1,L}\sim t_0^{-2}$ for odd $p$. In our case, we are specially
interested in the even $p$ case (2, 4 or 6 in Fig. \ref{fig:ed}) and
$t_0>1$. In all these cases we obtain a strong end-to-end correlation,
but the effect gets smaller for larger $p$.

We would like to remark that the $t_0^2$ dependence can be
heuristically justified assuming that the energy gap may be estimated
via Eq.~\eqref{eq:effcoup}, despite the fact that the Dasgupta-Ma SDRG
is not valid when the hoppings are not strongly inhomogeneous. In that
case, the energy scale associated to the edge state is given by
$t_0^2$.  Of course, a rigorous justification is still missing.

The relation between the end-to-end correlation and the energy gap in
the edge-dimerized chains is shown in the lower panel of
Fig.~\ref{fig:ed}. We see an interesting collapse of the values of $p$
by pairs: $\{1,2\}$, $\{3,4\}$ and $\{5,6\}$, i.e: cases $p=2m-1$ and
$p=2m$ provide the same results.

\begin{figure}
  \includegraphics[width=7.5cm]{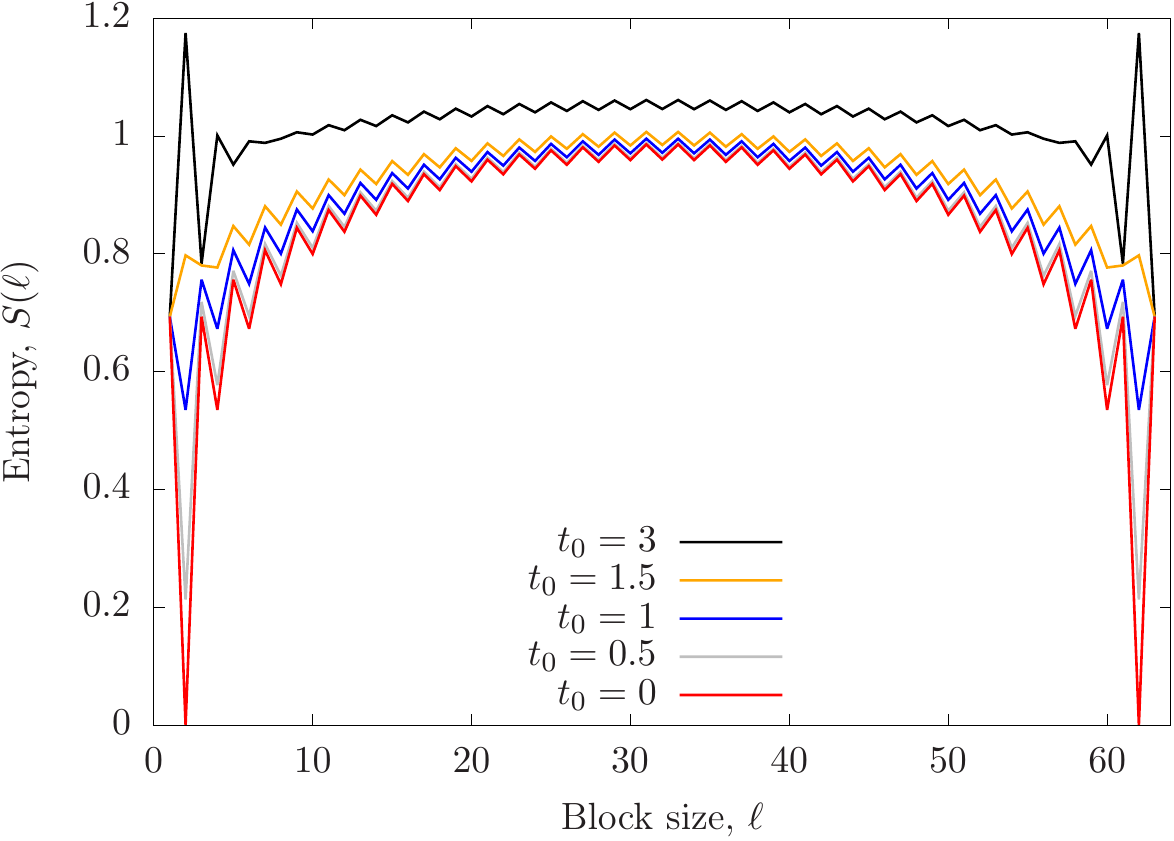}
  \includegraphics[width=7.5cm]{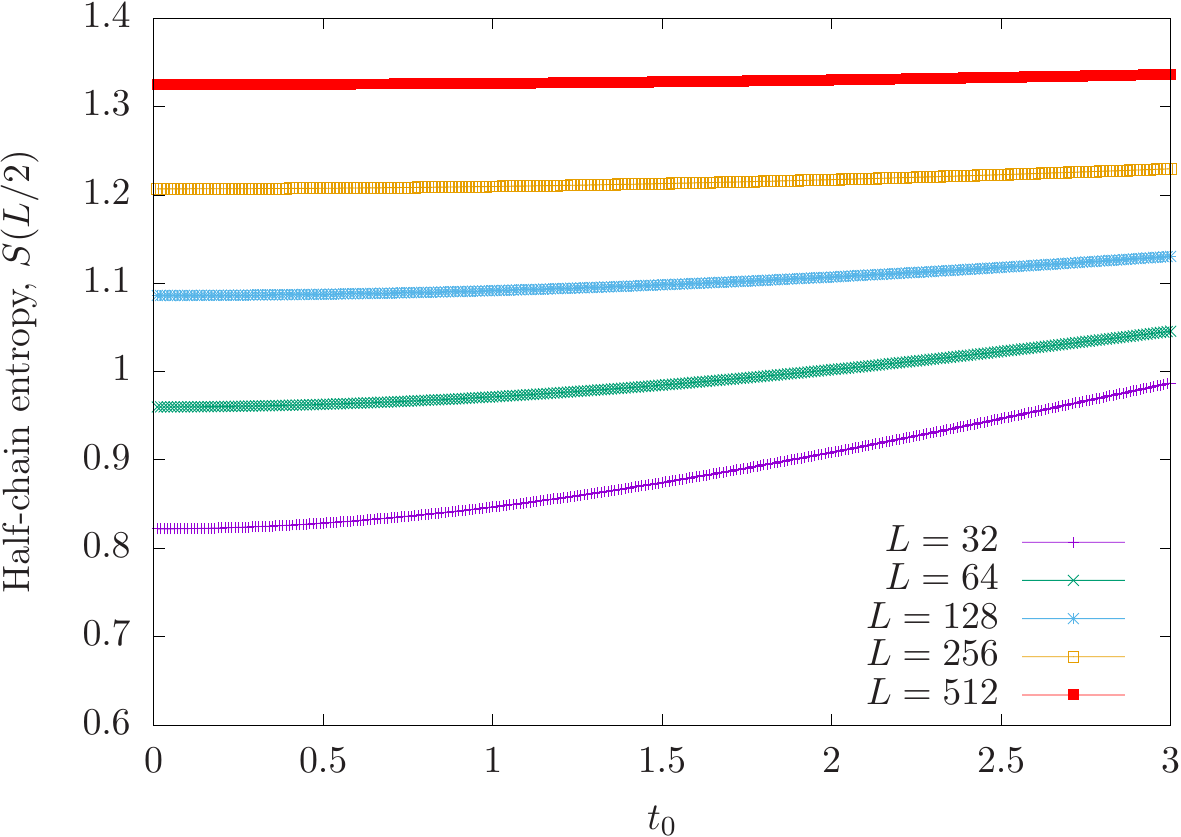}
  \includegraphics[width=7.5cm]{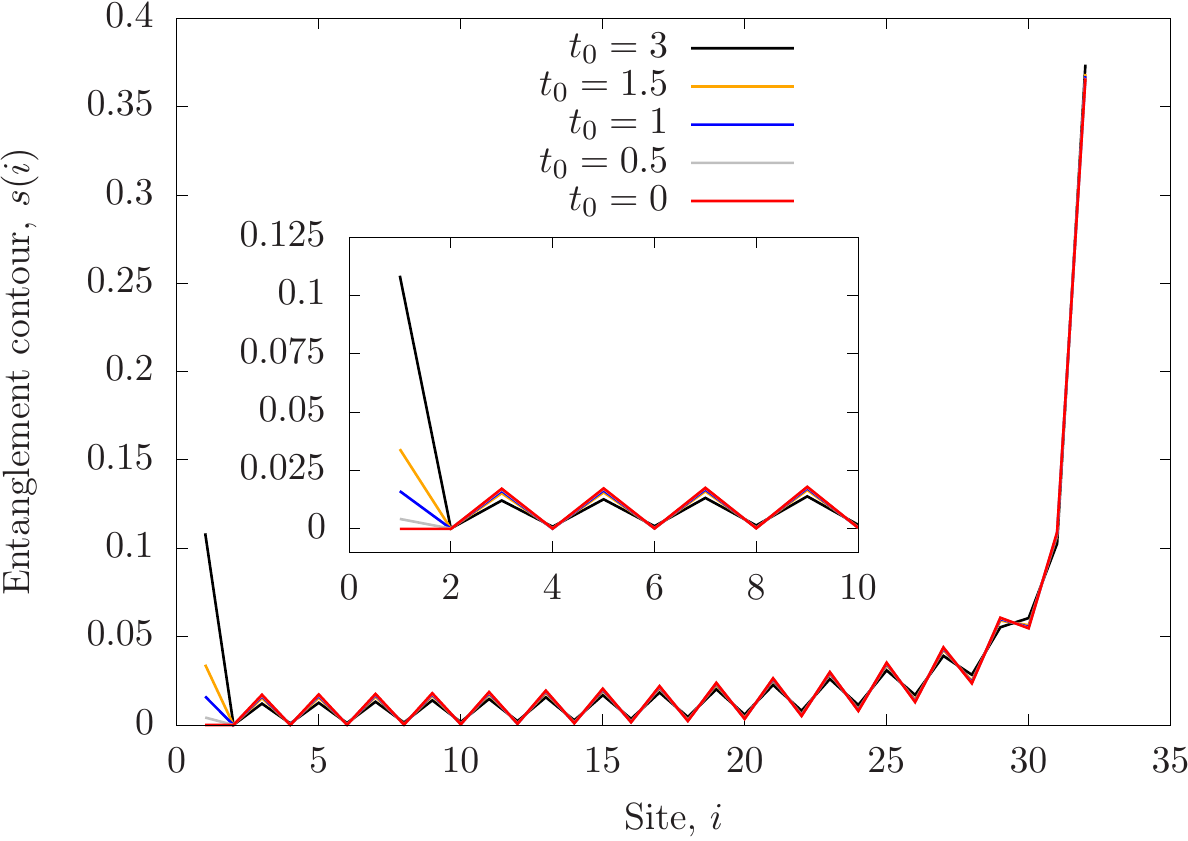}
  \caption{Entanglement structure of edge-dimerized states. Top:
    Entanglement entropy of blocks $\{1,\cdots,\ell\}$ as a function
    of the block size $\ell$ for different values of $t_0$, using
    $L=64$, on the GS of an edge-dimerized chain. Center: Half-system
    entanglement entropy $S(L/2)$ as a function of $t_0$, for
    different values of $L$. Notice the nearly flat behavior for large
    values of $L$. Bottom: entanglement contour for the left half of
    the $L=64$ system and the same values of $t_0$ used for the top
    panel. Inset: zoom on the left part of the block.}
  \label{fig:Sedge}
\end{figure}

Let us consider the entanglement structure of the edge-dimerized
chains. Fig. \ref{fig:Sedge} has a similar structure to that of
Fig. \ref{fig:entropy}. The top panel of Fig. \ref{fig:Sedge} shows
the entropy of blocks of the form $\{1,\cdots,\ell\}$ as a function of
$\ell$, $S(\ell)$, for a chain with $L=64$ and different values of
$t_0$. For $t_0=0$ the two extreme sites detach, while for $t_0=1$ we
obtain the clean result, predicted by CFT, Eq. \eqref{eq:Scft}. As we
increase $t_0$, the entropy becomes more flat, as in the bottom-right
panel of Fig. \ref{fig:engin}, and a high peak is obtained for
$S(2)$ and $S(L-2)$, denoting that the block $\{1,2\}$ cuts two bonds:
the long distance bond $(1,L)$ and the short distance bond
$(2,3)$. The central panel of Fig. \ref{fig:Sedge} shows the
half-chain entropy of the chain, $S(L/2)$ as a function of $t_0$ for
different values of $L$ which form a geometric progression. The
approximate arithmetic progression of the values denotes the
logarithmic dependence of the entropy with the system size, typical of
critical 1D systems, as opposed to the SSH case
(Fig. \ref{fig:entropy}, central). A relevant clue is provided by the
{\em entanglement contour} of the left-half of the system with $L=64$,
$s(i)$, plotted in the bottom panel of Fig. \ref{fig:Sedge} for
different values of $t_0$. Notice that the right extreme of the plot
corresponds to the center of the chain. The curves for all values of
$t_0$ are nearly identical, with a slight decrease of the bulk contour
for large $t_0$, while there is a strong increase in the contour of
first site, $s(1)$, which can be further noticed in the inset. This
implies that the main effect of the increase of $t_0$ is just to
create the edge state, the bond $(1,L)$, with a minimal distortion in
the rest of the entanglement structure.

We have obtained that these edge-dimerized chains, with strong hopping
amplitudes only in the second and penultimate links, give much better
results for the large end-to-end correlations {\em while} maintaining
a finite energy gap, a result that protects the edge state of the
system making it quite robust.

Let us provide some numerical comparison. Consider the SSH chain,
Eq. \eqref{eq:ham} with hoppings given in Eq. \eqref{eq:dimer}, using
$L=40$ and $\delta=0.2$ gives an end-to-end correlation of
$|C_{1,L}|\approx 0.3$, and an energy gap $\Delta E\approx 3.6\cdot
10^{-4}$. The same correlation can be obtained with an edge-dimerized
chain (very close to the optimal case) using $t_0\approx 6.35$, for
which the gap is now $\Delta E\approx 0.036$, i.e. 100 times
larger. For larger values of $\delta$, the gap ratio can be even
larger


\section{Conclusions and further work}
\label{sec:conclusions}

In this article we have explored 1D quantum systems in their ground
states which, despite their local interactions, can develop large
correlations between well separated sites (at the nanoscale). We have
only considered independent fermions, but heuristic arguments suggest
that similar structures could be found in presence of
interactions. E.g. the rainbow state can be obtained in presence of a
density-density repulsion \cite{Laguna.16}.

Open fermionic chains dimerize naturally in many relevant cases, due
to the Peierls instability, thus giving rise to a SSH Hamiltonian. If
the first and last hoppings are weak, a symmetry protected topological
state is formed, characterized by the presence of an edge state which
gives rise to very high end-to-end correlations. This edge-state can
be explained using entanglement monogamy: all bulk sites pair up,
leaving the first and last alone. Thus, a bond will be established
between them. Nonetheless, Dasgupta-Ma renormalization arguments show
that the energy gap associated with this state decreases faster than
exponentially, leading to very low stability under external
perturbations or a finite temperature.

We have developed a machine-learning algorithm in order to determine
the hopping pattern which can give rise to the maximal possible
end-to-end correlation for a given fixed energy gap on an open
fermionic chain. The results show that modulated dimerizations, which
are flat in the bulk, provide much better results. Optimality was
usually achieved by patterns which present strong hopping amplitudes
only in the second and penultimate links, which we have termed
edge-dimerized chains.

The differences in robustness between the GS of the SSH model and the
edge-dimerized one can be quite large: the energy gap can be more than
100 times larger for $L=40$ sites and a correlation of $0.3$ (being
$0.5$ the maximal value, for a Bell pair). The differences in the
entanglement structure are remarkable, and can help us understand the
enhanced stability of the edge-dimerized Hamiltonian. Indeed, the
entanglement entropy and contour show that the edge-dimerized GS is
virtually identical to the clean one in the bulk, with a huge
difference in the boundary. Thus, we can conjecture that the optimal
correlation is mainly obtained by leaving the entanglement structure
of the bulk untouched.

Of course, this stability can not be extended to arbitrarily large
chains, but it can be used to engineer nanoscopic quantum systems with
interesting properties comprising $\sim 50$-$100$ sites. Systems with
these types of hopping patterns can appear naturally in quantum wires
\cite{Ahn.03,Ahn.05} or organic molecules \cite{Gruner.88}, or can be
engineered using optical lattices using the so-called cold-atom
toolbox \cite{Lewenstein,toolbox}. On the other hand, spatial
modulations of the hoppings have been proposed to study the effects of
curved space-time on quantum matter and the Unruh effect
\cite{Celi.10,Laguna_Celi.17}.

This work constitutes a proof-of-principle that edge-dimerization can
help build strong long-distance correlations, along with some of the
phenomena associated. Further relevant work will consider the
applicability of these edge states for quantum information purposes,
possible condensed-matter realizations, extension to more dimensions
and dynamical effects.


\begin{acknowledgments}

We would like to acknowledge very useful discussions with
S.N. Santalla and G. Sierra. J.R.-L. acknowledges funding from the
Spanish Government through Grant No. FIS2015-69167-C2-1-P.

\end{acknowledgments}


\end{document}